# Thermally and Electrically Conductive Nanopapers from Reduced Graphene Oxide: effect of nanoflakes thermal annealing on the film structure and properties

M.M. Bernal[a], M. Tortello[b], S. Colonna[a], G.Saracco[b], A. Fina[a*]

[a] *Dipartimento di Scienza Applicata e Tecnologia, Politecnico di Torino, 15121 Alessandria, Italy*

[b] *Dipartimento di Scienza Applicata e Tecnologia, Politecnico di Torino, 10129 Torino, Italy*

[*]*Corresponding author: alberto.fina@polito.it*

**Abstract**

In this study, we report a different strategy to prepare graphene nanopapers from direct vacuum filtration. Instead of the conventional method, *i.e.* thermal annealing nanopapers at extremely high temperatures prepared from graphene oxide (GO) or partially reduced GO, we fabricate our graphene nanopapers directly from suspensions of fully reduced graphene oxide (RGO), obtained after RGO and thermally annealing at 1700 °C in vacuum. By using this approach, we studied the effect of thermal annealing on the physical properties of the macroscopic graphene-based papers. Indeed, we demonstrated that the enhancement of the thermal and electrical properties of graphene nanopapers prepared from annealed RGO is strongly influenced by the absence of oxygen functionalities and the morphology of the nanoflakes. Hence, our methodology can be considered as a valid alternative to the classical approach.

**1. Introduction**

Two-dimensional (2D) materials, *i.e.* one-atom thick layers of van der Waals materials, have gained worldwide attention because of their outstanding properties that arise from their structure and dimensionality.[1, 2] Graphene, a single-layer of $sp^2$ hybridized carbon atoms arranged in a honeycomb lattice, has aroused great interest as one of the most promising 2D materials. The unique physical properties of this carbon allotrope,[3-5] in particular its high electron mobility and ballistic conduction,[6-8] have suggested its use in many technological fields. However, real-world applications such as sensors, electronic devices, thermal management, energy storage conversion and EMI shielding to name only a few, often demand the development of flexible, lightweight, paper-like materials with high electrical and thermal conductivity and good corrosion resistance.[9-13] Thus, the fabrication of assembled architectures from graphene building blocks is of fundamental and practical significance to exploit the intrinsic features of individual graphene sheets at a macroscopic level.

Recent advances in the development of flexible graphene paper-like architectures with superior thermal and electrical properties have focused their attention on different assembly strategies.[9, 12-14] In particular, the flow-directed filtration-induced technique, where graphene suspensions are vacuum filtered, has attracted great interest for the manufacturing of free standing papers.[15-19] Hence, to achieve graphene films with superior properties, individual graphene sheets have to be previously dispersed and stabilized in a liquid medium while the electrostatic repulsive forces and bonding interactions need to be balanced to prevent their re-aggregation.[11, 15] A common approach has been the use of graphene oxide (GO) in water suspension as a precursor to obtain GO nanopapers with superior mechanical properties.[16, 20] The large interactions at the surface between nanosheets, the wrinkled structure morphology and the crosslinking of the functional groups are responsible for both the high Young's modulus and mechanical strength.[16, 18, 21, 22] Despite the oxygen-bearing groups facilitate the stabilization of GO in water or polar solvents, they disrupt the $sp^2$ hybridization of graphene layers, deteriorating the thermal and electrical conductivities of the nanoflakes and, in turn, reducing the performance of the as-prepared GO papers.

In order to restore the π-conjugated system of these materials, two different approaches have been reported. The first strategy is based on the reduction of the as-prepared GO films by chemical methods (using hydrazine hydrate or hydriodic acid), thermal processes or a combination of both. As a result of these processes, the conductivities are partially recovered due to the gradual removal of the oxygen species and water moisture between the nanosheets, packing them more tightly.[11, 17, 23, 24] Nevertheless, the mechanical properties and the structure integrity of the as-prepared GO or RGO films are seriously deteriorated after the thermal treatments, even at relatively low temperatures (~

200 °C).[23] The second approach consists in a previous chemical reduction of GO sheets using hydrazine hydrate to produce reduced graphene oxide (RGO) flakes, followed by sonication in organic solvents and vacuum filtration to produce RGO films.[17, 18, 22, 23, 25-28] In this case, the formation of RGO sheets implies the removal of many functional groups and thus, in order to achieve good dispersions, it is necessary to use organic solvents with surface tension similar to that of graphene such as dimethylformamide (DMF) or N-methyl pyrrolidone (NMP), the addition of surfactants to the aqueous medium or the formation of graphene colloids. Despite the interactions between oxygen-bearing groups are limited, the vacuum filtration-induced directional-flow allows the assembly of RGO sheets into well-ordered macroscopic structures with remarkable mechanical and electrical properties.[17, 18, 27, 29]

We have recently reported that the thermal properties of RGO flakes after thermal annealing at 1700 °C are increased due to the reduction of oxygen-functional groups and the ordering of the graphene structure.[30] This has encouraged us to manufacture conductive graphene nanopapers using an alternative strategy: direct vacuum filtration of both RGO flakes or thermally annealed RGO flakes (RGO_1700) suspensions. In this work, we present the successful fabrication of lightweight nanopapers from both conventional RGO and less defective thermally annealed RGO flakes as well as the assessment of their thermal and electrical properties.

## 2. Materials and Methods

*2.1. Fabrication of graphene nanopapers.*

RGO and RGO_1700 flakes, prepared as previously reported,[30] were suspended in DMF at concentrations of 0.15 mg mL$^{-1}$ and the solutions were sonicated in pulsed mode (15 sec on and 15 sec off) for 15 min with power set at 30 % of the full output power (750 W) by using an ultrasonication probe (Sonics Vibracell VCX-750, Sonics & Materials Inc.) with a 13 mm diameter Ti-alloy tip. The RGO and RGO_1700 suspensions were subjected to vacuum filtration using a Nylon Supported membrane (0.45 µm nominal pore size, diameter 47 mm, Whatman). After filtration, the as-obtained papers were peeled off from the membranes and dried at 65 °C under vacuum for 2 hours to completely remove the solvent. Finally, RGO and RGO_1700 nanopapers were mechanically pressed in a laboratory hydraulic press (Specac Atlas 15T) under a uniaxial compressive load of 5 kN for 10 min at 25 °C. The apparent density ($\rho_{app}$) of the samples was calculated according to the formula $\rho_{app} = m/V$, where $m$ is the mass of the nanopaper, weighed using a microbalance (Sensitivity: < 0.1 µg) and $V$ is calculated from a well-defined disk film using the average thicknesses measured as described in [23].

*2.2. Characterization.*

The morphology of the graphene papers was characterized by a high resolution Field Emission Scanning Electron Microscope (FESEM, ZEISS MERLIN 4248). The thermal diffusivity ($\alpha$) was measured using the xenon light flash technique (LFT) (Netzsch LFA 467 *Hyperflash*). The method is compliant with the international standard methods ASTM E-1461, ASTM E-2585, DIN 30905 and DIN EN-821. The samples were cut in disks of 23 mm and the measurements were carried out in a special in-plane sample holder. Each sample was measured five times at 25 °C. The in-plane thermal conductivity of the film, $\kappa$, was then calculated from the equation $\kappa=\rho\alpha C_p$, where $\rho$ is the apparent density of the graphene film ($\rho_{app}$) and $C_p$ is the specific heat capacity of graphite ($C_p = 0.71$ J (g K)$^{-1}$). In order to compare the thermal conductivity obtained on nanopapers with different density, owing to different porosity, the conductivity of dense nanoflakes networks were obtained by applying the Maxwell's effective medium approach. Therefore, the effective conductivity $\kappa$, is calculated from equation (1), where $\kappa_a$ is the apparent thermal conductivity of the nanopaper and $\kappa_{air}$ is the thermal conductivity of air.

$$\kappa_a = k \frac{\kappa_{air} + 2k + 2\varphi(k_{air} - k)}{\kappa_{air} + 2k - 2\varphi(k_{air} - k)} \tag{1}$$

The electrical conductivity was measured at room temperature by the standard four-point probe method. The electrical contacts were fabricated by means of a small drop of conductive silver paste and connected to an Agilent 34420A Nanovoltmeter. The conductivity was calculated by the formula $\sigma = \frac{d}{RS}$ where $R$ is the measured resistance, $d$ the distance between the voltage contacts and $S = tw$ where $t$ and $w$ are the thickness and width of the sample, respectively. The uncertainty on the conductivity was calculated by taking into account the uncertainties on the geometry of the sample. Electrical conductivity was also made independent of nanopaper porosity dividing the above calculated value by the volume fraction of conductive nanoparticles, accordingly with the method applied to thermal conductivity.

**3. Results and Discussions**

RGO and RGO_1700 were synthesized and characterized as described in our previous work.[30] RGO and RGO_1700 nanopapers were prepared by sonication of the nanoflakes in DMF for 15 min followed by vacuum filtration through a polyamide membrane (see the Materials and Methods Section for details). The resultant nanopapers were peeled off from the membrane, dried under vacuum and mechanically pressed. The fabrication of both graphene-based papers was carried out under the same temperature conditions, concentration of starting material in solution and applied force for compacting the films. The as-obtained free-standing nanopapers are in both cases highly flexible (Figure 1).

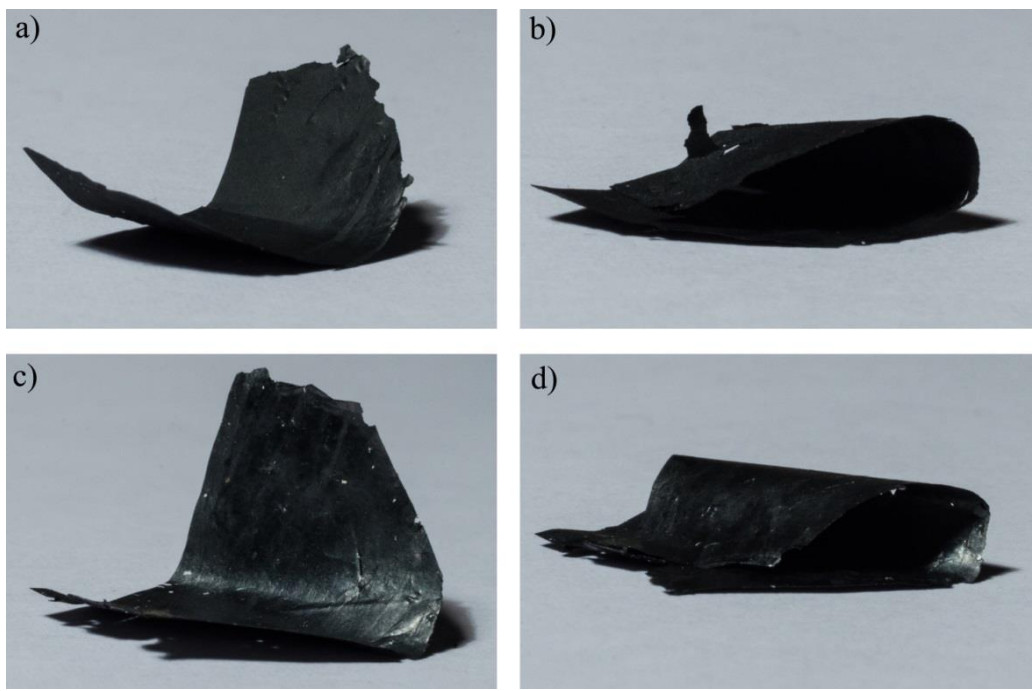

**Figure 1.** Photographs of free-standing nanopapers: (a) RGO nanopaper, (b) RGO nanopaper bent 180°, (c) RGO_1700 nanopaper and (d) RGO_1700 nanopaper bent 180°.

The cross-sectional FESEM images of the free-standing films (Figure 2) reveal a porous-like structure with micrometric cavities between the nanoflakes. The formation of irregularly stacked nanoflakes assemblies by vacuum filtration follows a semi-ordered accumulation mechanism.[31, 32] According to this process, the randomly oriented graphene flakes within a suspension form a loosely aggregated structure during the first stages of the filtration due to the weak adhesion between adjacent nanoflakes. The compressive forces generated as the solvent is removed, perpendicular to the direction of the solvent flow, produce a significant degree of order in the structure, even if this process does not completely align the resultant papers, as observed in the FESEM images. Furthermore, the crumpled structure of the initial RGO nanoflakes limits the packing of flakes in the nanopaper and cause the formation of microsized air cavities, resulting in a measured apparent density ($\rho_{app}$) of 0.4 g cm$^{-3}$. The RGO_1700 nanopaper (Figure 2 (c) and (d)) shows a more tightly packed and homogeneous structure compared to the RGO nanopaper (Figure 2 (a) and (b)). This difference may be attributed to both physical and chemical reasons. On the one hand, the thermal annealing is known to increase order within the stack of graphene layers in the thickness of a RGO nanoflake: indeed, annealing at high temperature was previously reported to increase correlation lengths both in plane and perpendicular to the plane of the nanoflakes.[30] Despite the overall morphology of the nanoflakes remains rather crumpled,[30] the increase in correlation length may result in a higher planarity at least in some regions of the nanoflakes, thus affecting the efficiency of nanoflakes stacking. On the other hand, the annealing treatment of the RGO flakes at 1700 °C reduces the number of oxidised groups on the nanoplatelets. The lack of significant oxidation reduces the steric hindrance arising from the residual functional groups, and enhances the possibility to build π-π interactions over extended areas, thus contributing in more efficient packing of flakes into the nanopaper,[33] finally leading to an apparent density of 1.0 g cm$^{-3}$ for RGO_1700 nanopapers.

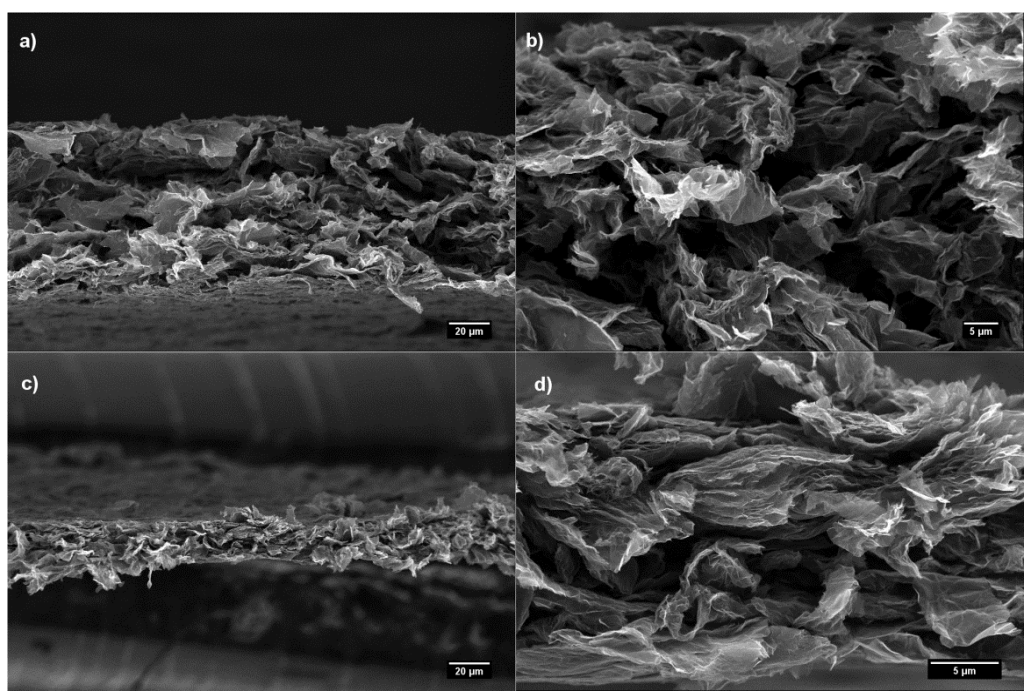

**Figure 2.** Cross-sectional FESEM images of (a) RGO nanopaper and (c) RGO_1700 nanopaper and corresponding higher magnifications (b) and (d), respectively.

Graphene-based nanopapers are microscopically heterogeneous materials which can be considered as graphene-air composites. Hence, the thermal and electrical conductivities of graphene-based nanopapers are dominated by their intrinsic structure which strongly depends not only on the intrinsic conductivity of the individual nanoflakes, but also on the size of the nanoplatelets, their orientation in the films, the porosity and the contact resistance.[34]

In plane thermal diffusivity for both RGO and RGO_1700 nanopapers was measured by xenon flash method, showing diffusivity values of 43.32 ± 1.51 and 198.56 ± 2.75 mm$^2$ s$^{-1}$ for pristine and annealed RGO, respectively. These values clearly evidence a better performance of the nanopaper obtained with annealed RGO nanoflakes. However, due to the significant differences in density and thickness between nanopapers prepared with pristine and annealed nanoflakes, such figures represent the performance in heat spreading on the different nanopapers (*i.e.* the manufacts) and not the intrinsic property of the dense network made of RGO/RGO_1700 nanoflakes. While an apparent in-plane thermal conductivity can simply be calculated from in-plane diffusivity measurements multiplied by the apparent nanopaper density and the heat capacity, a comparison between the intrinsic thermal conductivity of the dense networks made of RGO or RGO_1700 should be made by decoupling from the nanopaper density. Conductivity values of the nanoflake networks were therefore obtained by applying the Maxwell's effective medium approach, as detailed in the experimental section.

According to this calculations, in plane thermal conductivity of the dense nanoflakes networks were obtained at about 123± 4W m$^{-1}$K$^{-1}$ for the RGO nanopaper and 479± 7 W m$^{-1}$K$^{-1}$ for the RGO_1700 one, thus evidencing a dramatic increase of thermal transfer performance of the carbon nanoflakes network upon high-temperature annealing of the RGO nanoflakes. The enhancement of the thermal dissipation of the individual RGO flakes after thermal annealing, as a consequence of the removal of oxygen-functional groups with a simultaneous restoration of the *sp$^2$* hybridization of the C atoms, has already been proved by Scanning Thermal Microscopy (SThM), [30] even though a quantitative determination of the thermal conductivity from SThM measurements remains extremely challenging.[35] Here we demonstrate that the improvement of the thermal transfer efficiency for the annealed RGO flakes is directly correlated to a dramatic increase by magnitude factor of five at the macroscale. Beside the effect of annealing, it is worth noting that the thermal conductivity of the nanopaper prepared in this work by filtration of RGO is similar to that reported by Renteria *et al.*[9] for their dense rGO nanopaper prepared by deposition of GO, thermally reduced at 1000 °C, thus confirming the preparation method proposed in this work as a valid alternative to the conventional method. In fact, annealing rGO powder could routinely be carried out after conventional thermal reduction, and the obtained powders may be exploited for the preparation of nanopapers of large size, including by continuous processes, thus overcoming the limitation in terms of maximum size for the post treatment of the nanopapers in the traditional method.

We further studied the electrical conductivity of the RGO and RGO_1700 nanopapers using the four-point probe method. Electrical conductivities measured were normalized on RGO flakes volume fraction, accordingly with the method applied in thermal conductivity. The electrical conductivity increases from $(5.83 \pm 0.03) \times 10^3$ S m$^{-1}$ for the RGO flakes, to $(2.07 \pm 0.04) \times 10^4$ S m$^{-1}$ for the annealed ones. These results are in accordance with the improvements observed in the thermal properties. It is known that defects and incomplete reduction hinder the electrical performance of RGO, in terms of both reduced electrical conductivity and charge carrier mobility [36] and the thermal treatment favors the increase of sp$^2$ domains and thus the graphitization of the nanoflakes.[9] In the literature, GO nanopapers prepared in a similar way and treated at 800 °C[37] and 700 °C[25] showed electrical conductivities of $5.8 \times 10^3$ and $5.0 \times 10^3$ S m$^{-1}$, respectively, consistent with the values observed in this work for RGO papers, confirming the effectiveness of our preparation approach for nanopapers, as observed for the thermal properties.

## 4. Conclusions

In summary, our procedure to prepare graphene-based nanopapers by flow-directed filtration-induced technique using mild conditions and without post-treatments is effective to produce highly porous and flexible macroscopic all-graphene based structures. The structure of the resultant graphene nanopapers is strongly influenced by the defectiveness of the starting RGO flakes. The thermal and electrical properties were improved in nanopapers prepared with thermally treated RGO flakes, because the reduction of functional groups not only restores the $sp^2$ carbon structure but also generates the formation of higher density structures. The combination of the thermal and electrical properties accompanied by the low density of these macroscopic graphene-based materials makes these assembled architectures potential candidates in thermal management and electronic applications.

**Author's contribution**

A. Fina conceived the experiments, interpreted results and coordinated the project. M. Bernal carried out the preparation of nanopapers, performed most of the characterization/data analysis and participated to interpretation of results. M. Tortello carried out the electrical characterization/interpretation and contributed to interpretation of thermal conductivity . S. Colonna contributed to thermal characterization and interpretation. G. Saracco contributed to the discussion of the results. Manuscript was mainly written by M. Bernal and A. Fina.

**Acknowledgements**

This work has received funding from the European Research Council (ERC) under the European Union's Horizon 2020 research and innovation programme grant agreement 639495 — INTHERM — ERC-2014-STG. Funding from Graphene@PoliTo initiative of the Politecnico di Torino is also acknowledged.

The authors gratefully acknowledge Julio Gomez at AVANZARE Innovacion Tecnologica S.L. (E) for the provided RGO, Matteo Pavese for high temperature annealing of RGO, Mauro Raimondo for FESEM observations and Renato Gonnelli for the useful discussions.

**References**

1. Butler, S.Z., S.M. Hollen, L. Cao, Y. Cui, J.A. Gupta, H.R. Gutiérrez, T.F. Heinz, S.S. Hong, J. Huang, A.F. Ismach, E. Johnston-Halperin, M. Kuno, V.V. Plashnitsa, R.D. Robinson, R.S. Ruoff, S. Salahuddin, J. Shan, L. Shi, M.G. Spencer, M. Terrones, W. Windl, and J.E. Goldberger (2013) Progress, Challenges, and Opportunities in Two-Dimensional Materials Beyond Graphene. ACS Nano 7(4): 2898-2926
2. Sun, Z. and H. Chang (2014) Graphene and Graphene-like Two-Dimensional Materials in Photodetection: Mechanisms and Methodology. ACS Nano 8(5): 4133-4156
3. Balandin, A.A., S. Ghosh, W. Bao, I. Calizo, D. Teweldebrhan, F. Miao, and C.N. Lau (2008) Superior Thermal Conductivity of Single-Layer Graphene. Nano Letters 8(3): 902-907
4. Stoller, M.D., S. Park, Y. Zhu, J. An, and R.S. Ruoff (2008) Graphene-Based Ultracapacitors. Nano Letters 8(10): 3498-3502
5. Lee, C., X. Wei, J.W. Kysar, and J. Hone (2008) Measurement of the Elastic Properties and Intrinsic Strength of Monolayer Graphene. Science 321(5887): 385-388
6. Novoselov, K.S., A.K. Geim, S.V. Morozov, D. Jiang, Y. Zhang, S.V. Dubonos, I.V. Grigorieva, and A.A. Firsov (2004) Electric Field Effect in Atomically Thin Carbon Films. Science 306(5696): 666-669
7. Geim, A.K. and K.S. Novoselov (2007) The rise of graphene. Nat Mater 6(3): 183-191
8. Du, X., I. Skachko, A. Barker, and E.Y. Andrei (2008) Approaching ballistic transport in suspended graphene. Nat Nano 3(8): 491-495
9. Renteria, J.D., S. Ramirez, H. Malekpour, B. Alonso, A. Centeno, A. Zurutuza, A.I. Cocemasov, D.L. Nika, and A.A. Balandin (2015) Strongly Anisotropic Thermal Conductivity of Free-Standing Reduced Graphene Oxide Films Annealed at High Temperature. Advanced Functional Materials 25(29): 4664-4672
10. Hou, Z.-L., W.-L. Song, P. Wang, M.J. Meziani, C.Y. Kong, A. Anderson, H. Maimaiti, G.E. LeCroy, H. Qian, and Y.-P. Sun (2014) Flexible Graphene–Graphene Composites of Superior

Thermal and Electrical Transport Properties. ACS Applied Materials & Interfaces 6(17): 15026-15032

11. Huang, W., X. Ouyang, and L.J. Lee (2012) High-Performance Nanopapers Based on Benzenesulfonic Functionalized Graphenes. ACS Nano 6(11): 10178-10185

12. Xin, G., H. Sun, T. Hu, H.R. Fard, X. Sun, N. Koratkar, T. Borca-Tasciuc, and J. Lian (2014) Large-Area Freestanding Graphene Paper for Superior Thermal Management. Advanced Materials 26(26): 4521-4526

13. Shen, B., W. Zhai, and W. Zheng (2014) Ultrathin Flexible Graphene Film: An Excellent Thermal Conducting Material with Efficient EMI Shielding. Advanced Functional Materials 24(28): 4542-4548

14. Zhang, L., N.T. Alvarez, M. Zhang, M. Haase, R. Malik, D. Mast, and V. Shanov (2015) Preparation and characterization of graphene paper for electromagnetic interference shielding. Carbon 82: 353-359

15. Cong, H.-P., J.-F. Chen, and S.-H. Yu (2014) Graphene-based macroscopic assemblies and architectures: an emerging material system. Chemical Society Reviews 43(21): 7295-7325

16. Dikin, D.A., S. Stankovich, E.J. Zimney, R.D. Piner, G.H.B. Dommett, G. Evmenenko, S.T. Nguyen, and R.S. Ruoff (2007) Preparation and characterization of graphene oxide paper. Nature 448(7152): 457-460

17. Chen, H., M.B. Müller, K.J. Gilmore, G.G. Wallace, and D. Li (2008) Mechanically Strong, Electrically Conductive, and Biocompatible Graphene Paper. Advanced Materials 20(18): 3557-3561

18. Li, D., M.B. Muller, S. Gilje, R.B. Kaner, and G.G. Wallace (2008) Processable aqueous dispersions of graphene nanosheets. Nat Nano 3(2): 101-105

19. Yeh, C.-N., K. Raidongia, J. Shao, Q.-H. Yang, and J. Huang (2015) On the origin of the stability of graphene oxide membranes in water. Nat Chem 7(2): 166-170

20. Gong, T., D.V. Lam, R. Liu, S. Won, Y. Hwangbo, S. Kwon, J. Kim, K. Sun, J.-H. Kim, S.-M. Lee, and C. Lee (2015) Thickness Dependence of the Mechanical Properties of Free-Standing Graphene Oxide Papers. Advanced Functional Materials 25(24): 3756-3763

21. Dao, T.D., J.-E. Hong, K.-S. Ryu, and H.M. Jeong (2014) Super-tough functionalized graphene paper as a high-capacity anode for lithium ion batteries. Chemical Engineering Journal 250: 257-266

22. Kumar, P., F. Shahzad, S. Yu, S.M. Hong, Y.-H. Kim, and C.M. Koo (2015) Large-area reduced graphene oxide thin film with excellent thermal conductivity and electromagnetic interference shielding effectiveness. Carbon 94: 494-500

23. Paliotta, L., G. De Bellis, A. Tamburrano, F. Marra, A. Rinaldi, S.K. Balijepalli, S. Kaciulis, and M.S. Sarto (2015) Highly conductive multilayer-graphene paper as a flexible lightweight electromagnetic shield. Carbon 89: 260-271

24. Song, N.-J., C.-M. Chen, C. Lu, Z. Liu, Q.-Q. Kong, and R. Cai (2014) Thermally reduced graphene oxide films as flexible lateral heat spreaders. Journal of Materials Chemistry A 2(39): 16563-16568

25. Vallés, C., J. David Núñez, A.M. Benito, and W.K. Maser (2012) Flexible conductive graphene paper obtained by direct and gentle annealing of graphene oxide paper. Carbon 50(3): 835-844

26. Compton, O.C., D.A. Dikin, K.W. Putz, L.C. Brinson, and S.T. Nguyen (2010) Electrically Conductive "Alkylated" Graphene Paper via Chemical Reduction of Amine-Functionalized Graphene Oxide Paper. Advanced Materials 22(8): 892-896

27. Lin, X., X. Shen, Q. Zheng, N. Yousefi, L. Ye, Y.-W. Mai, and J.-K. Kim (2012) Fabrication of Highly-Aligned, Conductive, and Strong Graphene Papers Using Ultralarge Graphene Oxide Sheets. ACS Nano 6(12): 10708-10719

28. Chen, X., W. Li, D. Luo, M. Huang, X. Wu, Y. Huang, S.H. Lee, X. Chen, and R.S. Ruoff (2017) Controlling the Thickness of Thermally Expanded Films of Graphene Oxide. ACS Nano 11(1): 665-674

29. Gwon, H., H.-S. Kim, K.U. Lee, D.-H. Seo, Y.C. Park, Y.-S. Lee, B.T. Ahn, and K. Kang (2011) Flexible energy storage devices based on graphene paper. Energy & Environmental Science 4(4): 1277-1283

30. Tortello, M., S. Colonna, M. Bernal, J. Gomez, M. Pavese, C. Novara, F. Giorgis, M. Maggio, G. Guerra, G. Saracco, R.S. Gonnelli, and A. Fina (2016) Effect of thermal annealing on the heat transfer properties of reduced graphite oxide flakes: A nanoscale characterization via scanning thermal microscopy. Carbon 109: 390-401

31. Putz, K.W., O.C. Compton, C. Segar, Z. An, S.T. Nguyen, and L.C. Brinson (2011) Evolution of Order During Vacuum-Assisted Self-Assembly of Graphene Oxide Paper and Associated Polymer Nanocomposites. ACS Nano 5(8): 6601-6609

32. Sheath, P. and M. Majumder (2016) Flux accentuation and improved rejection in graphene-based filtration membranes produced by capillary-force-assisted self-assembly. Philosophical Transactions of the Royal Society A: Mathematical, Physical and Engineering Sciences 374(2060)

33. Shao, J.-J., W. Lv, and Q.-H. Yang (2014) Self-Assembly of Graphene Oxide at Interfaces. Advanced Materials 26(32): 5586-5612

34. Feng, W., M. Qin, and Y. Feng (2016) Toward highly thermally conductive all-carbon composites: Structure control. Carbon 109: 575-597
35. Gomès, S., A. Assy, and P.-O. Chapuis (2015) Scanning thermal microscopy: A review. physica status solidi (a) 212(3): 477-494
36. Bai, H., C. Li, and G. Shi (2011) Functional Composite Materials Based on Chemically Converted Graphene. Advanced Materials 23(9): 1089-1115
37. Zhang, L.L., X. Zhao, M.D. Stoller, Y. Zhu, H. Ji, S. Murali, Y. Wu, S. Perales, B. Clevenger, and R.S. Ruoff (2012) Highly Conductive and Porous Activated Reduced Graphene Oxide Films for High-Power Supercapacitors. Nano Letters 12(4): 1806-1812